\documentclass[prl,twocolumn,floatfix,preprintnumbers,amsmath,amssymb,superscriptaddress]{revtex4-1}

\usepackage{graphicx}
\usepackage{amsmath}
\usepackage{amssymb}
\usepackage{color}
\usepackage{dcolumn}
\usepackage{epsfig}
\usepackage{bm}
\usepackage{subfigure}
\usepackage[urlcolor=blue]{hyperref}
\hypersetup{backref, colorlinks=true, linkcolor=blue, citecolor=blue}

\begin{document}

\title{Synthesis of Superconducting Phase of La$_{0.5}$Ce$_{0.5}$H$_{10}$ at High Pressures}

\author{Ge Huang}
\affiliation{Key Laboratory of Materials Physics, Institute of Solid State Physics, Chinese Academy of Sciences, Hefei 230031, China}
\affiliation{University of Science and Technology of China, Hefei 230026, China}
\affiliation{Center for High Pressure Science and Technology Advanced Research, Shanghai 201203, China}

\author{Tao Luo}
\affiliation{School of Science, Harbin Institute of Technology, Shenzhen 518055, China}
\affiliation{Center for High Pressure Science and Technology Advanced Research, Shanghai 201203, China}

\author{Philip Dalladay-Simpson}
\affiliation{Center for High Pressure Science and Technology Advanced Research, Shanghai 201203, China}

\author{Liu-Cheng Chen}
\affiliation{School of Science, Harbin Institute of Technology, Shenzhen 518055, China}
\affiliation{Center for High Pressure Science and Technology Advanced Research, Shanghai 201203, China}

\author{Zi-Yu Cao}
\affiliation{Center for Quantum Materials and Superconductivity (CQMS) and Department of Physics, Sungkyunkwan University, Suwon 16419, Republic of Korea}
\affiliation{Center for High Pressure Science and Technology Advanced Research, Shanghai 201203, China}

\author{Di Peng}
\affiliation{Center for High Pressure Science and Technology Advanced Research, Shanghai 201203, China}

\author{Federico A. Gorelli}
\affiliation{Center for High Pressure Science and Technology Advanced Research, Shanghai 201203, China}
\affiliation{National Institute of Optics (INO-CNR) and European Laboratory for Non-Linear Spectroscopy (LENS), Via N. Carrara 1, 50019 Sesto Fiorentino (Florence), Italy}

\author{Guo-Hua Zhong}
\affiliation{Shenzhen Institute of Advanced Technology, Chinese Academy of Sciences, Shenzhen, 518055, Shenzhen, China}
\affiliation{University of Chinese Academy of Sciences, Beijing, 100049, China}

\author{Hai-Qing Lin}
\affiliation{Beijing Computational Science Research Center, Beijing, 100193, China}

\author{Xiao-Jia Chen}
\email{xjchen2@gmail.com}
\affiliation{Center for High Pressure Science and Technology Advanced Research, Shanghai 201203, China}
\affiliation{School of Science, Harbin Institute of Technology, Shenzhen 518055, China}

\date{\today}

\begin{abstract}

Clathrate hydride \emph{Fm}\={3}\emph{m}-LaH$_{10}$ has been proven as the most extraordinary superconductor with the critical temperature $T_c$ above 250 K upon compression of hundreds of GPa in recent years. A general hope is to reduce the stabilization pressure and maintain the high $T_c$ value of the specific phase in LaH$_{10}$. However, strong structural instability distorts \emph{Fm}\={3}\emph{m} structure and leads to a rapid decrease of $T_c$ at low pressures. Here, we investigate the phase stability and superconducting behaviors of \emph{Fm}\={3}\emph{m}-LaH$_{10}$ with enhanced chemical pre-compression through partly replacing La by Ce atoms from both experiments and calculations. For explicitly characterizing the synthesized hydride, we choose lanthanum-cerium alloy with stoichiometry composition of 1:1. X-ray diffraction and Raman scattering measurements reveal the stabilization of \emph{Fm}\={3}\emph{m}-La$_{0.5}$Ce$_{0.5}$H$_{10}$ in the pressure range of 140-160 GPa. Superconductivity with $T_c$ of 175$\pm$2 K at 155 GPa is confirmed with the observation of the zero-resistivity state and supported by the theoretical calculations. These findings provide applicability in the future explorations for a large variety of hydrogen-rich hydrides.

\end{abstract}

\maketitle
Metallic hydrides with a clathrate-like structure are proposed to possess high critical temperature $T_c$ particularly for the rare-earth hydrides LaH$_{10}$ and YH$_{10}$ \cite{wang12, li15, liu17, peng17}. Subsequently, a series of superhydride LaH$_{10 \pm x}$ was synthesised in succession by compressing initial La and H$_2$ upon laser heating at pressures up to 200 GPa \cite{geba18}. Superconductivity in LaH$_{10}$ is further evidenced with $T_c$ above 250 K from the resistance and magnetic susceptibility measurements \cite{somayazulu2019, drozdov2019, struz}. A theoretical work suggests that \emph{Fm}\={3}\emph{m} phase of LaH$_{10}$ can be stabilized by the quantum atomic fluctuations with the superconductivity at above 140 GPa \cite{errea}. Along this direction, various binary superhydrides have been extensively studied to facilitate $T_c$ by both the theoretical and experimental works \cite{chen20, li19, seme20, snid21}. Nevertheless, the necessity of extreme pressure conditions limits these high-\emph{T$_c$} superconductors for further examinations of the superconducting properties by various techniques. Finding a route to have a superconducting phase with a comparable $T_c$ at easily accessible pressures is a great challenge and on demand.

Among the known hydrides, \emph{Fm}\={3}\emph{m}-LaH$_{10}$ possesses superior superconducting behaviors with the maximum $T_c$ above 250 K. Considerable efforts are devoted to maintain the high symmetry phase of LaH$_{10}$ at lower pressures to preserve high $T_c$ values. However, owing to the strong instability, a structural distortion from \emph{Fm}\={3}\emph{m} to \emph{C}/2\emph{m} phase of LaH$_{10}$ leads to a steep drop of $T_c$ with decreasing pressure \cite{sun21}. Introducing extra electrons via metal doping into the hydrogen-rich materials is supposed to effectively reduce the metallization pressure \cite{sun19, zhang22, seme21}. The delocalized Ce 4\emph{f} electrons contribute to chemical pre-compression in cerium hydrides at pressure below 1 megabar with $T_c$ as high as 115 K \cite{chen20, jeon20}. Hence, one can expect low-pressure stability of high-$T_c$ superconducting \emph{Fm}\={3}\emph{m} phase of LaH$_{10}$ by introducing the Ce 4\emph{f} electrons. A recent experimental work demonstrates the synthesis of \emph{P}6$_3$/\emph{mmc} La$_{0.5}$Ce$_{0.5}$H$_{9}$ with $T_c$ up to 178 K \cite{bi22}. Concurrently, superconductivity with $T_c$ up to 190 K was reported in ternary polyhydrides \emph{hcp}-(La,Ce)H$_{9-10}$ and \emph{fcc}-(La,Ce)H$_{9-10}$\cite{chen22}. The uncertainties of the atomic ratio between La and Ce make difficult in performing the theoretical calculations in order to compare the results with experiments and thus hard to understand the behind microscopic mechanism of superconductivity in these La-Ce superhydrides.

In this work, we choose La$_{0.5}$Ce$_{0.5}$ alloy as the precursor material. The phase behaviors of the phonon vibrations and structural properties are examined through both the Raman spectroscopy and x-ray diffraction (XRD) techniques. The electrical transport measurements are carried out to detect the superconductivity through the realization of the zero-resistance state of the synthesized samples. The electronic structure and phonon spectrum together with the $T_{c}$ value are examined from the theoretical calculations. The comparison of the obtained $T_{c}$ with the early works is made to show the advantage of alloying hydrides to maintain high $T_{c}$ in the famous LaH$_{10}$ structure but at the moderate pressure.

\begin{figure}[t!]
  \includegraphics[width=\columnwidth]{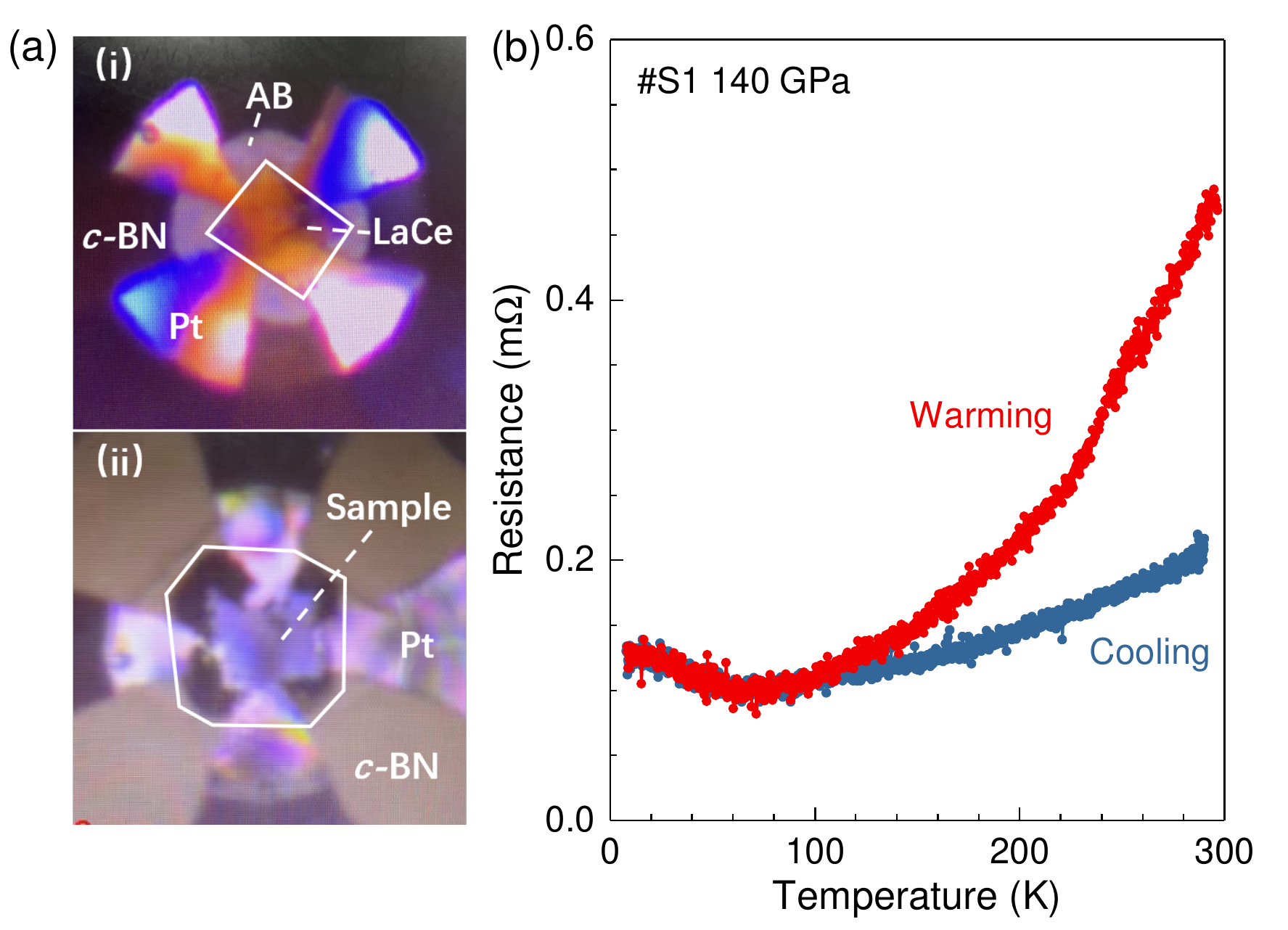}
  \caption{(a) Photographs of sample S1 before (i) and after (ii) laser heating. (b) Temperature-dependent resistance.}
\end{figure}

The type-Ia diamonds with 80 $\mu$m culets single beveled to 300 $\mu$m were used for all the measurements. The ammonia borane (AB) was used as the hydrogen source owing to the decomposition reaction upon heating. The La$_{0.5}$Ce$_{0.5}$ flake and AB were substantially loaded into diamond anvil cells (DAC) inside the argon glovebox (O$_2$, H$_2$O $\leqslant$ 0.1 ppm). Raman spectra of the first-order diamond edge were used to determine the pressure \cite{akah10}. Pure Ce (99.8 \%, Alfa Aesar) and La (99.9 \%, Alfa Aesar) were weighted and melted more than six times using an arc-melting machine under a high purity argon atmosphere. Powder XRD data were collected at Shanghai Synchrotron Radiation Facility and intergraded into two dimensional data set using the Dioptas software \cite{pres15}. The data was further analyzed using the LeBail method \cite{bail05} based on the Jana2006 program \cite{jana06}. Raman scattering measurements were performed in back scattering geometry with a laser wavelength of 488 nm.

The density functional theory calculations were carried out by using Vienna ab initio simulation package \cite{kres93, kres96, bloc94}. The exchange-correlation energy function was determined based on the Perdew-Burke-Ernzerhof form of the generalized gradient approximation \cite{pbe96}. The structures were optimized with high accuracy. The plane-wave cutoff energy was set as 600 eV. QUANTUM ESPRESSO package \cite{qe09} was used to calculate the phonon dispersion and the electron phonon coupling (EPC) constants with the 2$\times$2$\times$2 \emph{q}-points and 8$\times$8$\times$8 \emph{k}-points. The Allen-Dynes-modified McMillan equation \cite{allen75} was used to calculate $T_c$ in the following form: $T_c$ = $\frac{\omega_{log}}{1.2}\exp[-\frac{1.04(1+\lambda)}{\lambda-\mu^\ast(1+0.62\lambda)}]$, with $\omega_{log}$ = $\exp[\frac{2}{\lambda}\int\ln(\omega)\frac{\alpha^2F(\omega)}{\omega}d\omega]$ and $\lambda$ = 2$\int\frac{\alpha^2F(\omega)}{\omega}d\omega$, where $\mu^\ast$, $\alpha^2F(\omega)$, and $\lambda$ are the Coulomb pseudopotential, the electron phonon spectral function, and the EPC parameter, respectively.

\begin{figure}[t!]
  \includegraphics[width=\columnwidth]{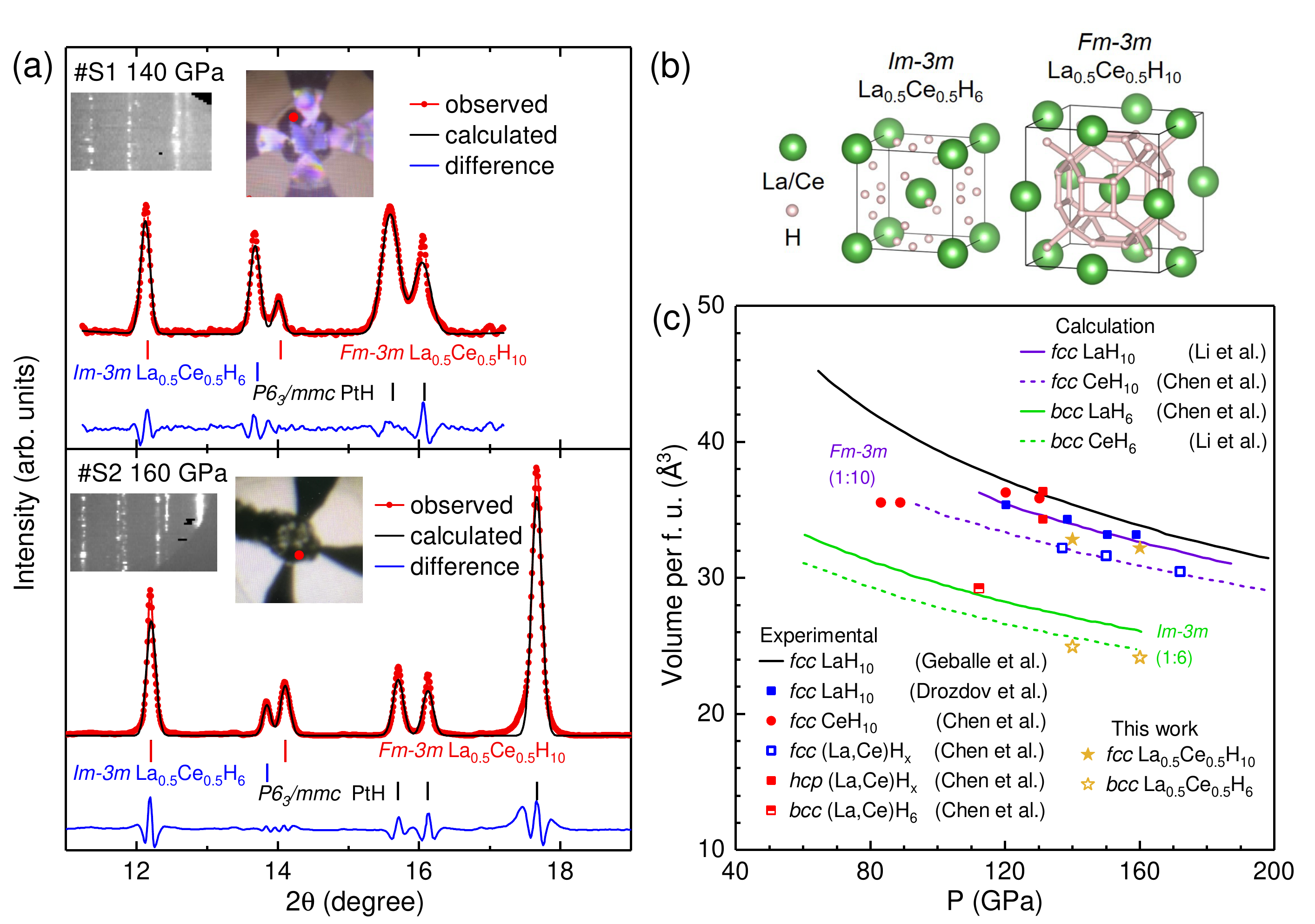}
  \caption{(a) Representative XRD patterns ($\lambda$ = 0.6199 {\AA}) at the edge of the samples marked by the red points in the inset photographs. Refinement results are plotted together with the measured XRD profiles. The vertical sticks at the lower patterns represent the reflection markers for the synthesized phases. The difference curves are given at the bottom of each panel. The intergraded XRD patterns are shown in the insets. (b) Crystal structures of La$_{0.5}$Ce$_{0.5}$H$_{10}$ and La$_{0.5}$Ce$_{0.5}$H$_{6}$. (c) Pressure-dependent cell volume per formula unit. The experimental data points and the theoretical results from Geballe $et$ $al$. \cite{geba18}, Drozdov $et$ $al$. \cite{drozdov2019}, Chen $et$ $al$. \cite{chen20, chen22} and Li $et$ $al$. \cite{li19} are plotted together for the comparison.}
\end{figure}

To synthesize the potential superconducting ternary La-Ce hydrides, we prepared three samples, labelled as S1, S2 and S3. In the first experimental run, the sample S1 is compressed to 140 GPa and then transferred for laser heating. Figure 1(a) shows the optical micrographs of the sample with the attached four Pt leads on the insulating \emph{c}-BN gasket inside the DAC before and after the laser heating. The size and color of the sample change significantly, indicating the incorporating of La$_{0.5}$Ce$_{0.5}$ alloy and the released H$_2$ upon heat treatment. Figure 1(b) shows the resistance as a function of temperature from 10 to 300 K for the heated sample S1. At low temperatures, the resistance minimum is clearly observed. The occurrence of the resistivity minimum has been extensively studied as the Kondo effect in the cerium metals and the yttrium- or lanthanum- based alloys containing the cerium impurities \cite{gey71, diet72}. Such anomalous behaviors of the Ce metals can be attributed to the presence of an 4\emph{f} level close to the Fermi level, thus producing a large resonant-scattering effect \cite{corn72}. In addition, the morphology at the center of the sample remains almost the same, suggesting the possible incomplete reaction of the alloys with hydrogen. It seems likely that La$_{0.5}$Ce$_{0.5}$ alloy dominates the electrical transport of sample S1.

\begin{table}[t!]
\begin{center}
\caption{Indexed lattice parameters and the unit cell volumes ($V$) for the formed phases in the samples S1 and S2.}
\label{table:1}
\begin{tabular}{m{2.2cm}<{\centering}  m{3.1cm}<{\centering}  m{3.1cm}<{\centering}}
\toprule\noalign{\smallskip}      & S1 (140 GPa) & S2 (160 GPa) \\
\noalign{\smallskip}\hline\noalign{\smallskip}   La$_{0.5}$Ce$_{0.5}$H$_{10}$ \ (\emph{Fm}\={3}\emph{m})  &  $a$=5.083$\pm$0.002 {\AA}, \ $V$=131.3$\pm$0.2 {\AA}$^3$  &  $a$=5.050$\pm$0.001 {\AA}, \ $V$=128.79$\pm$0.08 {\AA}$^3$  \\
\noalign{\smallskip}\hline\noalign{\smallskip}   La$_{0.5}$Ce$_{0.5}$H$_{6}$ \ (\emph{Im}\={3}\emph{m})  & $a$=3.681$\pm$0.002 {\AA}, \ $V$=49.88$\pm$0.08 {\AA}$^3$ & $a$=3.637$\pm$0.003 {\AA}, \ $V$=48.1$\pm$0.1 {\AA}$^3$  \\
\noalign{\smallskip}\hline\noalign{\smallskip}   PtH \ (\emph{P}6$_3$/\emph{mmc})  & $a$=2.638$\pm$0.002 {\AA}, \ $c$=4.442$\pm$0.002 {\AA}, \ $V$=26.77$\pm$0.04 {\AA}$^3$ &  $a$=2.619$\pm$0.001 {\AA}, \ $c$=4.419$\pm$0.001 {\AA}, \ $V$=26.25$\pm$0.02 {\AA}$^3$  \\
\noalign{\smallskip} \hline \hline
\end{tabular}
\end{center}
\end{table}

Since the further pressurization favors to stabilize the notable reaction product, the pressure in DAC for S2 was further increased to 160 GPa. To identify the crystal structures of the synthesized hydride, we thus conducted the \emph{in-situ} synchrotron XRD experiments for samples S1 and S2. Figure 2(a) shows the experimental XRD patterns obtained at the edge of the samples. By indexing the observed peaks, we can identify the appearance of three phases in the samples with \emph{Fm}\={3}\emph{m}, \emph{Im}\={3}\emph{m} and \emph{P}6$_3$/\emph{mmc} structures. The cell parameters for the formed phases are obtained as listed in Table I. Even though the accurate occupancy of the H atoms can not be determined owing to the weak x-ray scattering cross-section, the stoichiometry of synthesized hydrides can be obtained through the cell unit volume.

Figure 2(c) shows the \emph{P}-\emph{V} data points together with early experimental and theoretical works. For \emph{Fm}\={3}\emph{m}-La$_{0.5}$Ce$_{0.5}$H$_{10}$, the volumes per formula unit with 32.83 and 32.20 {\AA}$^3$ are comparable to those of \emph{Fm}\={3}\emph{m}-LaH$_{10}$ and CeH$_{10}$ at 140 and 160 GPa, respectively. The ratio of the hydrogen to the metal atoms is 10 to 1 in the synthesized hydride with \emph{Fm}\={3}\emph{m} structure, similar to the binary hydrides LaH$_{10}$ and CeH$_{10}$. Therefore, it can also be inferred that the metal atom sublattice is randomly occupied by the La and Ce atoms in \emph{Fm}\={3}\emph{m}-La$_{0.5}$Ce$_{0.5}$H$_{10}$. Simultaneously, the formed \emph{bcc} phase can be assigned as \emph{Im}\={3}\emph{m}-La$_{0.5}$Ce$_{0.5}$H$_{6}$, consistent to the early work \cite{chen22}. Binary \emph{Im}\={3}\emph{m} phases of LaH$_6$ and CeH$_6$ have not been experimentally synthesized yet. Moreover, theoretical predictions indicate the thermodynamic instability of \emph{Im}\={3}\emph{m}-LaH$_{6}$ \cite{liu17}. The introduction of 4\emph{f} electrons is beneficial to stabilize the \emph{Im}\={3}\emph{m} structure in LaH$_6$. We find obvious reflection peaks from the reaction product of the hydrogen and the Pt electrode, indicating the formation of the hexagonal PtH hydride. The \emph{P}6$_3$/\emph{mmc} phase of PtH has been synthesized earlier \cite{sche11}. Figure 2(b) shows the crystal structures of \emph{bcc} La$_{0.5}$Ce$_{0.5}$H$_{6}$ and \emph{fcc} La$_{0.5}$Ce$_{0.5}$H$_{10}$, respectively.

\begin{figure}[t!]
  \includegraphics[width=\columnwidth]{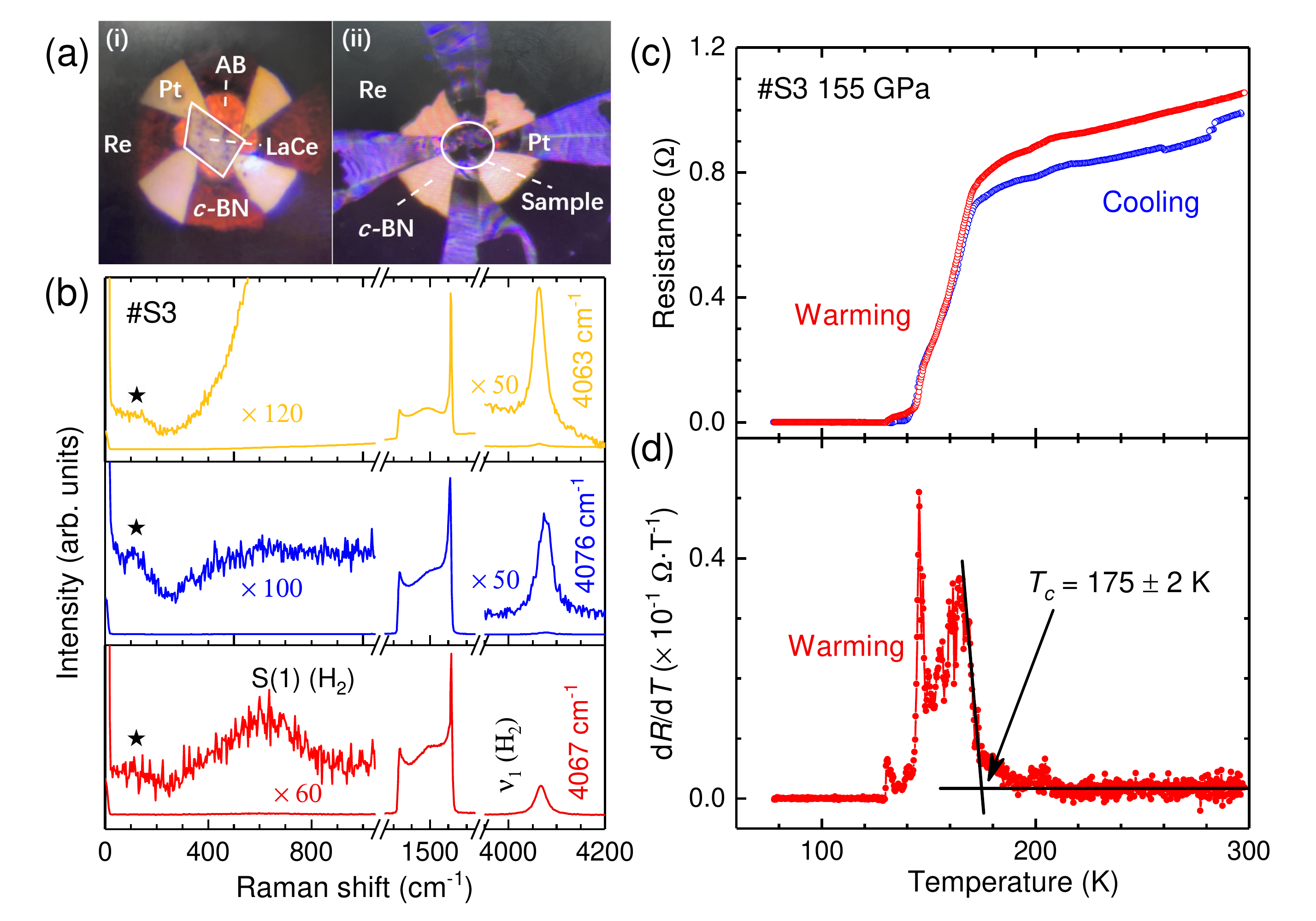}
  \caption{(a) Optical microscopy inside the DAC for the sample S3 before (i) and after (ii) laser heating at pressure of 155 GPa. (b) Raman spectra for the La-Ce-H system at different areas in the sample chamber. The blue and yellow curves show the Raman spectra obtained at the center of the sample from different side, the red curve shows the Raman spectra at the edge of the sample. The spectra are scaled for clarity. (c) Temperature-dependent resistance for sample S3. (d) Temperature-dependent d\emph{R}/d\emph{T} gives a \emph{T$_c$} value of 175$\pm$2 K.}
\end{figure}

According to the experimental XRD results, we find the stabilization of \emph{Fm}\={3}\emph{m}-La$_{0.5}$Ce$_{0.5}$H$_{10}$ superhydride in the pressure range of 140-160 GPa. To examine the superconductivity in the ternary La-Ce-H system, we immediately prepare DAC for S3 for the resistivity measurements. Figure 3(a) shows the photographs of the sample S3 before and after laser heating at 155 GPa. We can also observe the striking expansion of the sample. Raman spectroscopy is effective to detect the atomic structures through probing the vibrational modes. Figure 3(b) shows the Raman spectra of sample S3 after laser heating. The S(1) rotational band and the $\nu_1$ vibration mode at around 620 and 4270 cm$^{-1}$ indicate the existence of hydrogen inside the whole sample chamber \cite{heme90}. The variations in the frequencies of the $\nu_1$ mode indicate the slightly anisotropic conditions inside DAC. The excess of the released hydrogen further confirms the formation of the \emph{Fm}\={3}\emph{m} La$_{0.5}$Ce$_{0.5}$H$_{10}$ phase. The small peak at around 120 cm$^{-1}$ is probably due to the vibrations of the heavy elements of La and Ce of the synthesized superhydride phases. This is consistent with the early calculated phonon spectra at around the pressure studied for either LaH$_{10}$ \cite{liu17,peng17} or CeH$_{10}$ \cite{lib19}.

Figure 3(c) shows the temperature-dependent resistance of sample S3 in the warming and cooling cycles. A sharp drop of resistance below 180 K is clearly observed, suggesting the occurrence of superconductivity within the ternary La-Ce-H system. The zero-resistance state is achieved at lower temperatures. The broad superconducting transition is attributed to the various anisotropic stress inside the DAC. The $T_c$ value of 175$\pm$2 K is defined from the temperature derivative of the resistance (d\emph{R}/d\emph{T}) in the warming cycle as shown in Fig. 3(d). The obtained $T_c$ of 175$\pm$2 K in present sample shows consistency to the $T_c$ values in the recently discovered \emph{P}6$_3$/\emph{mmc}-La$_{0.5}$Ce$_{0.5}$H$_{9}$, \emph{P}6$_3$/\emph{mmc}-(La,Ce)H$_{9-10}$ and \emph{Fm}\={3}\emph{m}-(La,Ce)H$_{9-10}$ \cite{bi22, chen22}. In addition, PtH with \emph{hcp} structure exhibits superconductivity with a $T_c$ of 7 K at 30 GPa \cite{mats19}. However, $T_c$ of PtH was reported to decrease upon compression and becomes undetectable at the pressure range. Meanwhile, \emph{Im}\={3}\emph{m}-LaH$_6$ was reported to stabilize kinetically with an estimated $T_c$ of 176 K at 180 GPa \cite{seme21}, the superconducting order may be suppressed for the strongly correlated interactions of 4\emph{f} electrons. Therefore, we attribute the present superconducting phase to \emph{Fm}\={3}\emph{m}-La$_{0.5}$Ce$_{0.5}$H$_{10}$.

\begin{figure}[t!]
  \includegraphics[width=\columnwidth]{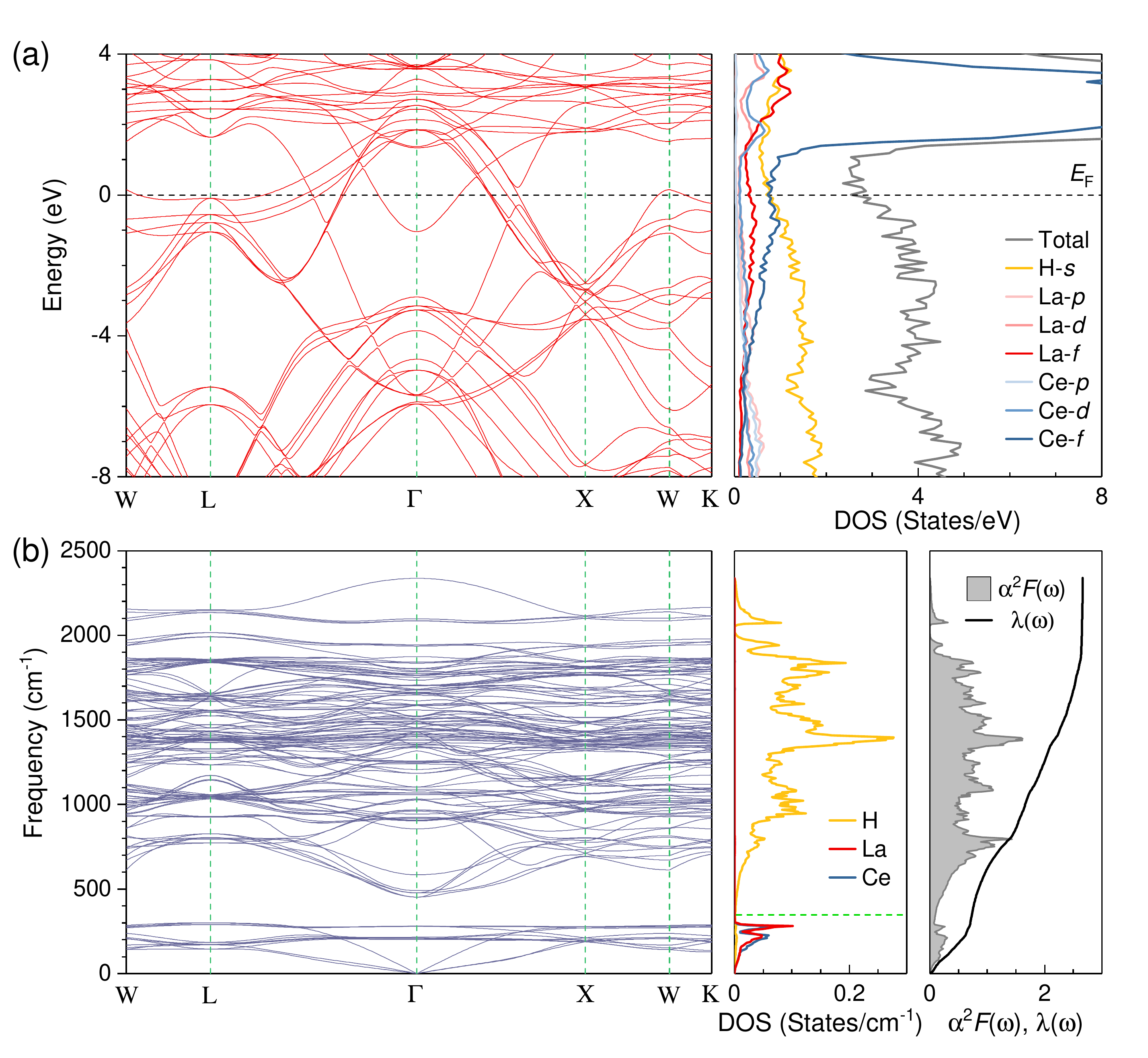}
  \caption{Electronic structure and phonon dispersion of La$_{0.5}$Ce$_{0.5}$H$_{10}$ at 300 GPa. (a) Calculated band structure, total and partial electronic DOS. The zero energy represents the Fermi level. (b) The phonon spectrum, projected phonon DOS, Eliashberg spectral function $\alpha^2F(\omega)$ and EPC integration $\lambda(\omega)$. The horizon line is drawn as the guide to the eyes.}
\end{figure}

The rare earth ratio of 1:1 for the alloy allows us to evaluate the electronic and phonon contributions to the observed superconductivity in \emph{fcc} La$_{0.5}$Ce$_{0.5}$H$_{10}$ in terms of the theoretical calculations. For the consideration of the structural stability as well as the nearly constant $T_{c}$ behavior of the \emph{fcc} LaH$_{10}$ phase \cite{somayazulu2019,drozdov2019,struz} at high pressures, we took the pressure of 300 GPa in the calculations. Figure 4(a) shows the electronic band structures, the total and partial electron states of density (DOS). It has been illustrated that the electronic DOS at the Fermi level [$N(E_F$)] in LaH$_{10}$ are dominated by the strong hybridization of the La 4\emph{f} and H \emph{s} orbitals \cite{wang19}. However, for La$_{0.5}$Ce$_{0.5}$H$_{10}$, the \emph{f} electrons of Ce and the \emph{s} electrons of H are the main compositions to $N(E_F$). Therefore, the introduction of the Ce elements in the studied hydrides is mainly to reduce the phase stability pressure in terms of the chemical pre-compression from the \emph{f} electrons. However, the observed superconductivity largely results from the H \emph{s} electrons.  In such alloy hydrides, the contributions to superconductivity from the La elements seem negligible.

\begin{figure}[t!]
  \includegraphics[width=\columnwidth]{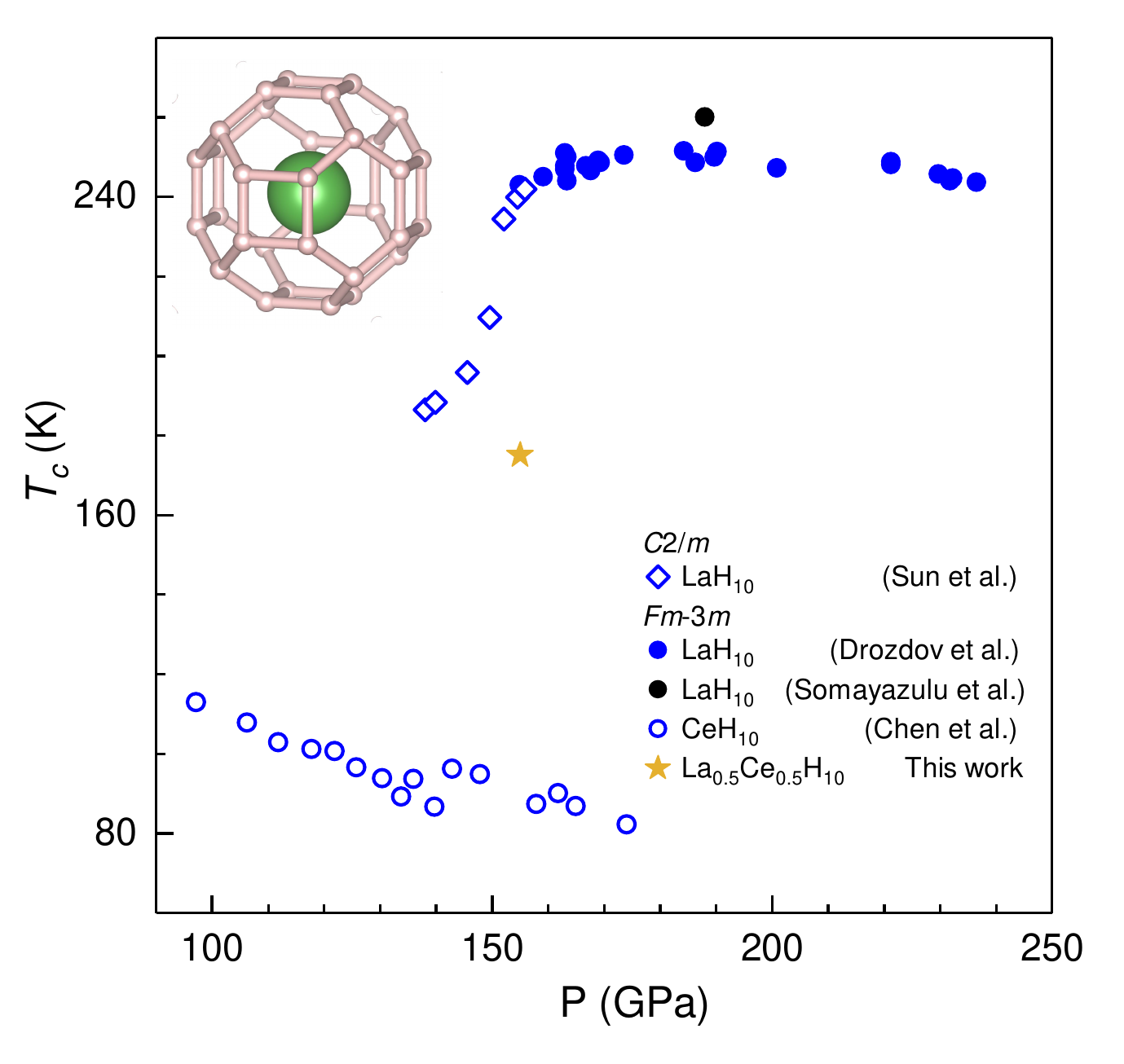}
  \caption{Pressure dependence of $T_c$ for superhydrides LaH$_{10}$, CeH$_{10}$ and La$_{0.5}$Ce$_{0.5}$H$_{10}$. The pressure of the data points from Drozdov $et$ $al.$ \cite{drozdov2019} and Sun $et$ $al.$ \cite{sun21} is increased by 18 GPa owing to the pressure scale difference between the hydrogen vibration and the diamond Raman edge. Inset shows the metal centered H$_{32}$ cage in the \emph{Fm}\={3}\emph{m} structure.}
\end{figure}

The phonon dispersion curves, the phonon DOS projected onto selected atoms, the Eliashberg function $\alpha^2F(\omega)$, and the intergraded EPC constant $\lambda(\omega)$ were calculated as shown in Fig. 4(b). The absence of imaginary frequency along the high symmetry \emph{k} path indicates the thermodynamic stability of \emph{Fm}\={3}\emph{m}-La$_{0.5}$Ce$_{0.5}$H$_{10}$. We note that the phonon spectrum can be divided into two regions. The lower branches with frequencies below 350 cm$^{-1}$ are mainly contributed to the motions of La and Ce atoms, in good agreement with the obtained Raman spectra [Fig. 3(b)]. The higher frequencies branches are dominated by hydrogen vibrations. An EPC parameter $\lambda$ of 2.69 is obtained from the combined contributions of $\sim$ 28\% and 72\% for the low and high branches, respectively. The large EPC is derived from the high-frequency optical modes of hydrogen. Compared to the $\lambda$ value of 0.69 and 1.86 in \emph{fcc} CeH$_{10}$ and LaH$_{10}$ at same conditions \cite{lib19, wang19}, we find the strengthened EPC in La$_{0.5}$Ce$_{0.5}$H$_{10}$ owing to the replacement of La by Ce atoms, strongly supporting our motivation. Using the Eliashberg equation \cite{elia60}, we calculated $T_c$ of 164-182 K for La$_{0.5}$Ce$_{0.5}$H$_{10}$ by using the typical values of 0.10-0.13 for the Coulomb potential $\mu^\ast$. The experimentally observed $T_c$ of 175$\pm$2 K is well reproduced by the calculations, indicating that the electron-phonon interactions can account for the superconductivity of \emph{Fm}\={3}\emph{m}-La$_{0.5}$Ce$_{0.5}$H$_{10}$.

The dependence of $T_c$ on pressure for the present ternary La$_{0.5}$Ce$_{0.5}$H$_{10}$ and the binary LaH$_{10}$ and CeH$_{10}$ from earlier works \cite{somayazulu2019, drozdov2019, chen20, sun21} is summarized in Fig. 5. $T_c$ of La$_{0.5}$Ce$_{0.5}$H$_{10}$ just lies at the middle points of the $T_c^{\prime}$s of the respective binary hydrides. Meanwhile, the \emph{Fm}\={3}\emph{m} La$_{0.5}$Ce$_{0.5}$H$_{10}$ can be dynamically stabilized in the studied pressure range of 140-160 GPa. However, a monoclinic structural distortion leads to the \emph{C}2/\emph{m} phase in the binary LaH$_{10}$ hydrides below the pressure of 160 GPa \cite{sun21}. Therefore, it can be appraised that the 4\emph{f} electrons in Ce metal are effective to reduce the stabilization pressure of the phases with high hydrogen content of 10.

In summary, we have synthesized the superconducting hydride La$_{0.5}$Ce$_{0.5}$H$_{10}$ with $T_c$ of 175 K at 155 GPa using the substantial alloy La$_{0.5}$Ce$_{0.5}$ as a starting material. Compared to the binary LaH$_{10}$, we found the dynamical stabilization of \emph{Fm}\={3}\emph{m} structure in La$_{0.5}$Ce$_{0.5}$H$_{10}$ at relatively low pressures. The delocalized 4\emph{f} electronic states in the Ce metal are proposed to produce the enhanced chemical pre-compression in the ternary La-Ce-H system. The phase stability and the obtained $T_{c}$ value find the supports from the theoretical calculations. These results open up an opportunity to search for the high-temperature superconductors at low or even ambient pressure in the solid solution alloy superhydrides.

\begin{acknowledgments}

The work at HIT was supported by the Basic Research Program of Shenzhen (Grant No. JCYJ20200109112810241) and the Shenzhen Science and Technology Program (Grant No. KQTD20200820113045081). The work at HPSTAR was supported by the National Key R\&D Program of China (Grant no. 2018YFA0305900).

\end{acknowledgments}

\end{document}